\documentclass[letter,twocolumn]{jpsj2} 
\usepackage{bm}
\usepackage{color}

\title{Magnetic Order in the Spin-1/2 Kagome Antiferromagnet Vesignieite}

\author{Makoto \textsc{Yoshida}\thanks{E-mail address: yopida@issp.u-tokyo.ac.jp}, 
Yoshihiko \textsc{Okamoto}, 
Masashi \textsc{Takigawa}, 
and Zenji \textsc{Hiroi}}

\inst{Institute for Solid State Physics, University of Tokyo, Kashiwa, Chiba 277-8581, Japan}

\abst{We report results of magnetic susceptibility and $^{51}$V- and $^{63,65}$Cu-NMR 
measurements on a high-quality powder sample of vesignieite BaCu$_3$V$_2$O$_8$(OH)$_2$, 
a candidate for the spin-1/2 kagome antiferromagnet. We observed a two-step 
magnetic transition: the appearance of spatially inhomogeneous static moments below 13 K and a 
long-range order below 9 K. The NMR data indicate a ${\bf Q}$ = 0 magnetic structure at 1.4 K 
with the in-plane spin components of three sublattices oriented at nearly 120$^{\circ}$ to each 
other and the magnitude of the ordered moments of at least 0.6~$\mu_B$.} 

\kword{BaCu$_3$V$_2$O$_8$(OH)$_2$, vesignieite, quantum spin system, kagome lattice, NMR, magnetic order}

\begin{document}
\maketitle

Understanding the ground state of the spin-1/2 antiferromagnetic Heisenberg model on a kagome lattice, 
a two-dimensional network of corner-sharing equilateral triangles, is a challenge in current 
condensed matter physics. Theories have proposed various ground states, such as spin liquids 
with no broken symmetry or valence-bond-crystal states~\cite{Waldtmann,Hermele,Singh}. 
Experimental efforts on quantum spin 1/2 systems have focused on materials, such as 
Cu$_3$V$_2$O$_7$(OH)$_2 \cdot $2H$_2$O (volborthite)~\cite{Hiroi,HYoshida,Yoshida1,Yoshida2,YoshidaHF}, 
BaCu$_3$V$_2$O$_8$(OH)$_2$ (vesignieite)~\cite{Okamoto,Colman,Quilliam,HYoshida2}, or  
ZnCu$_3$(OH)$_6$Cl$_2$ (herbertsmithite)~\cite{Shores,Helton}. 
These materials, however, deviate from the ideal kagome model owing to disorder, structural distortion, 
Dzyaloshinsky-Moriya (DM) interaction, or longer range interactions. Some of these effects
have been theoretically investigated~\cite{Cepas,Schnyder,Wang,Domenge}. 
In classical kagome systems, a long-range order with the $\sqrt{3} \times  \sqrt{3}$ pattern
is favored by the order-by-disorder effect \cite{Reimers}, while the DM 
interaction stabilizes the ${\bf Q}$ = 0 structure~\cite{Elhajal}. 

The crystal structure of vesignieite has been reported to be monoclinic with the space group $C2/m$.~\cite{Zhesheng} 
In this structure, vesignieite has slightly distorted kagome layers formed by isosceles triangles with two Cu sites and two 
kinds of exchange interaction $J$ and $J'$ (Fig.~1)~\cite{Okamoto}. 
The difference between the Cu1-Cu2 and Cu2-Cu2 bond lengths is very small at 0.2\%,~\cite{Okamoto} 
according to ref.~\citen{Zhesheng}. 
Examination of various Cu-O bond lengths points to 
the $d(3z^2 - r^2)$ orbitals for the unpaired $d$-electron at both Cu sites, which approximately satisfy 
the three-fold symmetry, as shown in Fig.~1~\cite{Okamoto}. 
Therefore, $J$ and $J'$ are expected to be nearly equal. 
Recently, the structure has been reexamined by Colman \textit{et al.}, who reported a 
bond length difference of 0.07\%~\cite{Colman}, which is negligibly small within the experimental error. 
More recently, H. Yoshida \textit{et al.} have reported a trigonal structure on the 
average with random occupation at the V and O sites.~\cite{HYoshida2} 
Although there seems to be a small discrepancy between these reports,\cite{Zhesheng,Colman,HYoshida2} 
vesignieite should be a promising candidate to explore properties of isotropic kagome systems. 
The magnetic susceptibility $\chi$ obeys the Curie-Weiss law $\chi  = C/(T + \theta _W)$ above 200 K 
with $\theta _W$ = 77 K.~\cite{Okamoto} A spin liquid-like ground state was initially suggested because the specific heat 
and $\chi $ data did not indicate any sign of magnetic order down to 2 K~\cite{Okamoto}. 
More recently, however, Colman \textit{et al.} have reported splitting of 
the field-cooled (FC) and zero-field-cooled (ZFC) magnetization data below 9~K~\cite{Colman}. 
Muon spin relaxation and nuclear magnetic resonance (NMR) studies also suggested an unusual ground 
state, where small frozen moments of 0.1-0.2~$\mu_B$ coexist with dynamically fluctuating spins~\cite{Colman, Quilliam}. 

\begin{figure}[b]
\begin{center}
\includegraphics[width=0.6\linewidth]{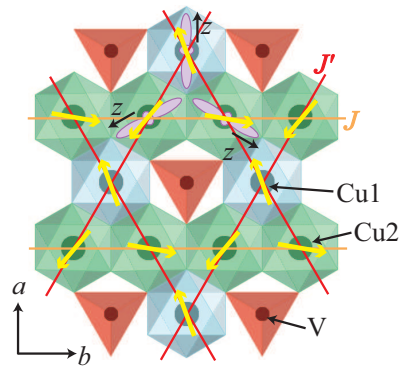}
\end{center}
\caption{(Color online) Schematic structure of vesignieite projected onto the $a$-$b$ plane. 
The H and O sites are not shown. The V sites are located below and above the Cu kagome layers, 
which are related by inversion with respect to the Cu sites. Hence, all V sites are equivalent. 
The arrows on the Cu sites schematically illustrate 
a 120$^{\circ }$ spin structure with the ${\bf Q}$ = 0 propagation vector. The ellipsoids represent the $d(3z^2 - r^2)$ orbital.}
\label{fig1}
\end{figure}

In this letter, we report results of $\chi $ and NMR measurements on a powder sample of 
vesignieite with a much improved quality. We observed a static internal field at the $^{51}$V nuclei 
below 13 K and a long-range magnetic order below 9~K. In contrast to the previous studies~\cite{Colman,Quilliam}, 
our NMR results indicate that essentially all spins are involved in a ${\bf Q}$ = 0 magnetic order with the 
in-plane moments on the three sublattices oriented at nearly 120$^{\circ}$ with each other. The ordered moments 
are estimated to be at least 0.6 $\mu_B$. The magnetic order is likely to be stabilized by a sufficiently large DM interaction.  

Powder samples of vesignieite were synthesized, as described in ref.~\citen{Okamoto}. 
They were then annealed under hydrothermal condition at 580 $^{\circ }$C and 60 MPa for 24 h. 
A significant improvement in crystalline domain size was achieved by hydrothermal annealing, 
as indicated by much shaper powder X-ray diffraction (XRD) peaks [Fig.~2(a)] than those in the 
previous samples~\cite{Okamoto}. A SQUID magnetometer (Quantum Design MPMS)
was used to measure $\chi$. The $^{51}$V-NMR spectra were obtained by summing the Fourier 
transform of the spin-echo signal obtained at equally spaced rf frequencies with a fixed 
magnetic field $B$, using the pulse sequence $\pi /2 - \tau  - \pi /2$. 
The $^{63,65}$Cu-NMR spectrum at zero field was obtained by recording the 
integrated intensity of the spin-echo signal at discrete frequencies. 
We determined 1/$T_1$ by fitting the spin-echo intensity $M(t)$ as a function of the time $t$ 
after several saturating comb pulses to the stretched exponential recovery function 
\begin{equation}\label{eq:stretch}
M(t)=M_{\rm eq}-M_0 \exp \left\{-(t/T_1)^{\beta} \right\}.
\end{equation}
This functional form was used to quantify the inhomogeneous distribution of 1/$T_1$
with the stretch exponent $\beta$, which is close to one for homogeneous relaxation.

The XRD pattern shows additional peaks from a small amount of 
secondary phases, as indicated by the arrows in Fig. 2(a). In NMR measurements, 
signals from such phases, even if observable, should appear at different 
frequencies from the main signal. Since no signal of $^{51}$V from secondary phases was observed in this 
sample, our NMR results represent properties of the main phase.  

\begin{figure}[tb]
\begin{center}
\includegraphics[width=0.8\linewidth]{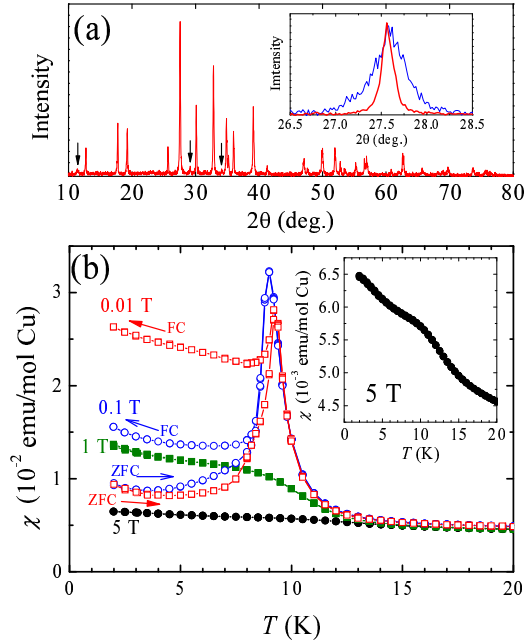}
\end{center}
\caption{(Color online) (a) Powder XRD pattern at room temperature. The inset shows 
a peak near 28 deg. of the annealed (red thick line) and nonannealed (blue thin line) samples. 
(b) $T$ dependence of $\chi $ at various magnetic fields. A clear difference between 
the FC and ZFC magnetizations is observed at 0.01 and 0.1 T, while they overlap each other at 1 and 5 T. 
The data at 5 T are expanded in the inset. 
}
\label{fig2}
\end{figure}

Figure 2(b) shows the temperature ($T$) dependence of $\chi$ at various magnetic fields $B$. 
A significant increase in $\chi$ is observed below 13~K for $B$ less than 1~T.  In addition, a sharp 
peak is observed at 9~K, below which deviation develops between the FC and ZFC conditions, indicating 
a weak ferromagnetism. A similar but less pronounced anomaly was reported previously~\cite{Colman}. 
With increasing $B$, the peak and hysteresis in $\chi$
become suppressed. However, the magnetic transition appears to persist up to 5 T, 
as indicated by a kink in $\chi$ shown in the inset of Fig. 2(b). 

Figure 3 shows the $T$ dependence of the $^{51}$V-NMR spectra at 3~T. The spectrum at 30~K shows a 
powder pattern expected for a paramagnetic phase consisting of a sharp center line 
and a quadrupole-broadened satellite line. In the higher $T$ range between 100 and 300 K, 
the magnetic hyperfine shift $K$ and $\chi$ follow the same $T$ dependence, yielding the hyperfine 
coupling constant $A_{\mathrm{hf}}^{\mathrm{V}}$ = 0.87~T/$\mu _B$. This is slightly larger 
than the value reported by Quilliam \textit{et al.} (0.77~T/$\mu _B$)~\cite{Quilliam}. 

\begin{figure}[tb]
\begin{center}
\includegraphics[width=0.7\linewidth]{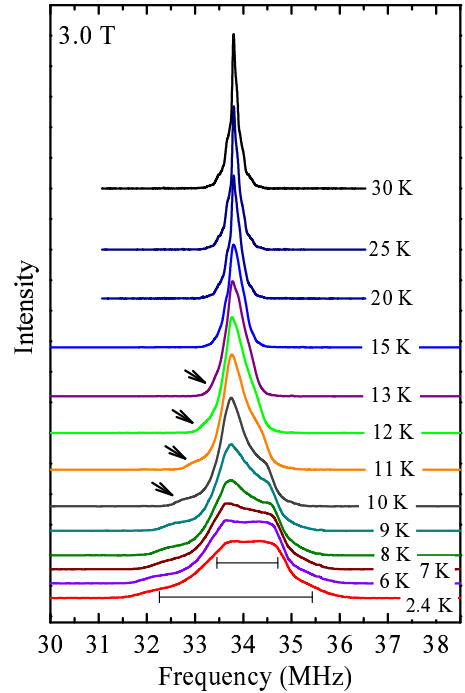}
\end{center}
\caption{(Color online) $T$ dependence of the $^{51}$V-NMR spectra at $B$ = 3.0~T}
\label{fig3}
\end{figure}

In the lower $T$ range, the NMR line shape depends strongly on $T$. 
Below 13~K, the spectra show a sudden asymmetric broadening with a shoulder, which shifts 
to lower frequencies with decreasing $T$, as indicated by the arrows in Fig.~3.
Such a line shape, which cannot be explained by either the anisotropic shift in the paramagnetic phase 
nor the spontaneous internal field with a fixed magnitude, 
indicates the appearance of internal fields due to spatially inhomogeneous static moments. 
With a further decrease in $T$, the line shape changes to the sum of two rectangular 
spectra, a narrow one (0.12~T wide) on top of a broader one (0.29~T wide), as indicated by the 
upper and lower bars in Fig.~3, respectively. 
Since a simple antiferromagnetic order leads to a rectangular NMR 
powder pattern~\cite{YoshidaHF}, this suggests a relatively simple spin structure at low temperatures.
Both the narrow and broad spectra show a tail extending to lower frequencies, but 
the reason for this is not well understood yet. In contrast to the previous observation~\cite{Quilliam}, 
we did not find an appreciable loss of NMR intensity at low temperatures. 
Therefore, we conclude that all Cu spins are essentially involved in the long-range ordering.   

As an appropriate measure of the line width of the spectra, we have examined the second moment 
defined as $M_2 = \int (\nu  - ~^{51}\gamma M_1)^2I(\nu )d\nu/(^{51}\gamma)^2$, 
where $^{51}\gamma$ = 11.1988~MHz/T is the nuclear gyromagnetic ratio, 
$I(\nu)$ is the NMR spectrum as a function of the frequency $\nu$ normalized as $\int I(\nu )d\nu  = 1$, 
and $M_1 = \int \nu I(\nu )d\nu /^{51}\gamma$ is the center of gravity of the spectrum. 
The inset of Fig.~4 shows the $T$ dependence of $\sqrt{M_2}$ at 3~T. A sudden increase in 
$\sqrt{M_2}$ is clearly observed below 13 K. It tends to saturate below 5 K.

\begin{figure}[tb]
\begin{center}
\includegraphics[width=0.8\linewidth]{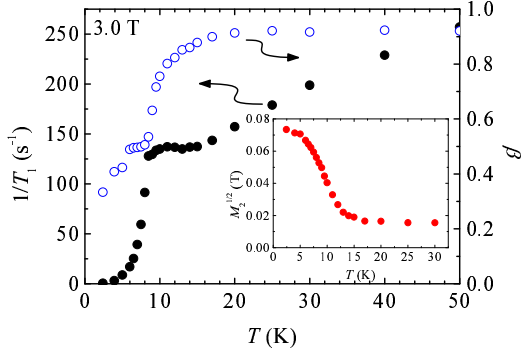}
\end{center}
\caption{(Color online) $T$ dependences of 1/$T_1$ (closed cirles) and the stretch exponent $\beta$ (open circles) 
for $^{51}$V nuclei at $B$ = 3.0~ T. The inset shows the $T$ dependence of $\sqrt{M_2}$ at 3~T.}
\label{fig4}
\end{figure}

Figure 4 shows the $T$ dependences of 1/$T_1$ and $\beta$ (eq.~\ref{eq:stretch}) at the $^{51}$V sites
for $B$ = 3.0~T. Above 20 K, $\beta$ is larger than 0.9, indicating a homogeneous nuclear relaxation. 
With decreasing $T$ below 15~K, 1/$T_1$ begins to 
show an inhomogeneous distribution. This is likely to be associated with the 
appearance of inhomogeneous static moments mentioned above. A sudden and steep decrease in 
1/$T_1$ occurs below 8.5~K, whereas $\chi$ shows a peak at 9~K, further supporting 
a phase transition into a long-range magnetic order. (The slight difference in 
transition temperature may be due to a finite field of 3~T in the 1/$T_1$ measurements.) 
Combining the $T$ dependences of $\chi$, $\sqrt{M_2}$, and 1/$T_1$, 
we conclude that the magnetic transition takes place in two steps. 
First, inhomogeneous static moments appear near 13~K, as indicated by the increase in 
line width (inset of Fig. 4) accompanied by the inhomogeneous distribution of 1/$T_1$ (Fig. 4) 
and the increase in $\chi$ [Fig. 2(b)]. This is then followed by a long-range order at 
$T_N$ = 9~K. The absence of a sharp peak in 1/$T_1$ at $T_N$, which is usually observed in 
antiferromagnets owing to the critical slowing down of the spin fluctuations, is an anomalous feature 
also reported previously~\cite{Quilliam}. This may be explained by 
the inhomogeneous development of static moments above $T_N$. 

To obtain information about the spin structure, we have investigated 
the $^{63,65}$Cu-NMR spectrum at zero magnetic field and  at $T$ = 1.4~K (Fig.~5).  
Two large broad peaks at 87 and 130 MHz and small peaks near 47 MHz are observed. 
We were not able to obtain reliable data below 40~MHz owing to a heavy acoustic ringing and a low 
signal intensity. This spectrum provides direct evidence for a magnetic order 
with a large moment. 

\begin{figure}[tb]
\begin{center}
\includegraphics[width=0.7\linewidth]{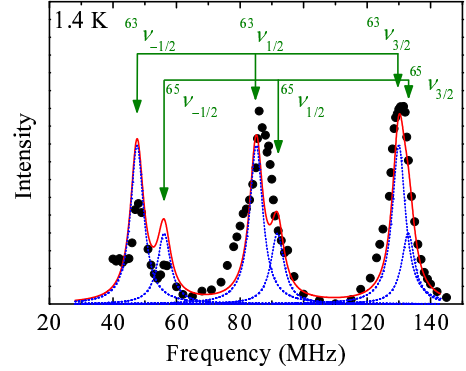}
\end{center}
\caption{(Color online) The zero-field $^{63,65}$Cu-NMR spectrum at $T$ = 1.4 K. 
The solid line is a calculated spectrum for a single value of $B_{\mathrm{int}}$ consisting 
of six peaks shown by the dotted lines. }
\label{fig5}
\end{figure}

A spontaneous ordered moment ${\bf M}$ generates the internal field 
${\bf B}_{\mathrm{int}} = {\bf A}_{\mathrm{hf}}^{\mathrm{Cu}}\cdot {\bf M}$ of several 
T/$\mu _B$ on Cu nuclei and gives rise to a zero-field resonance, 
where ${\bf A}_{\mathrm{hf}}^{\mathrm{Cu}}$ is the on-site hyperfine coupling tensor. 
Since both $^{63}$Cu and $^{65}$Cu nuclei have spin $I$ = 3/2, the interaction between the 
nuclear quadrupole moments $^{\alpha}Q$ ($\alpha$ = 63 or 65) and the electric field gradient 
(EFG) splits the zero-field resonance into three lines~\cite{Abragam}. 
The frequency of the center line is given by 
$^{\alpha}\gamma B_{\mathrm{int}}$, while the split satellite lines appear at 
$^{\alpha}\gamma B_{\mathrm{int}} \pm~^{\alpha}\nu_Q|3\mathrm{cos}^2\theta - 1 + \eta \mathrm{cos}2\phi \mathrm{sin}^2\theta |/2$ 
up to the first-order perturbation in the quadrupole interaction~\cite{Abragam}. Here, $^{\alpha}\gamma$ is the nuclear 
gyromagnetic ratio of $^{\alpha}$Cu nuclei, $\theta$ and $\phi$ specify the direction of
${\bf B}_{\mathrm{int}}$ with respect to the principal axes of EFG, $\eta$ is a parameter indicating 
the deviation of EFG from axial symmetry, and $^{\alpha}\nu_Q$ 
is the maximum quadrupole shift proportional to $^{\alpha}Q$. 
Since both isotopes have similar values of $\gamma$ and $Q$, $^{65}\gamma /^{63}\gamma = 1.0712$ and 
$^{65}Q/^{63}Q = 0.9252$, the two sets of quadrupole split three lines nearly overlap. Similar zero-field 
Cu-spectra have been observed in a number of antiferromagnets.~\cite{YoshidaCuBr}
The solid line in Fig.~5 is an example of the calculated spectrum with one set of parameters 
($B_{\mathrm{int}}$ = 7.7~T, $^{63}\nu _Q$ = 45~MHz, $\theta$ = 12~deg, and $\eta$ = 0). 
Although the value of $B_{\mathrm{int}}$ can be directly determined from the center peaks, 
an independent determination of $^{\alpha }\nu _Q$ and $\theta$ requires the examination of 2nd and higher order terms
and involves large uncertainties. 

The well-resolved zero-field Cu-NMR spectrum indicates that the magnitude of 
${\bf B}_{\mathrm{int}}$ and its direction with respect to the principal axes of EFG 
are nearly the same for all Cu sites. If there were more than two Cu sites with largely different 
values of $B_{\mathrm{int}}$, the center line must have split or significantly broadened. Also, 
the different directions of ${\bf B}_{\mathrm{int}}$ must have split or broadened the satellite lines. 
As already mentioned, we expect the three-fold rotational symmetry of an isotropic kagome lattice 
to be approximately preserved in vesignieite because the distortion is very small (Fig.~1). Therefore, 
the same symmetry is expected for EFG and ${\bf B}_{\mathrm{int}}$ at all Cu sites. 
The uniqueness of ${\bf B}_{\mathrm{int}}$ then leads us to conclude that the structure of 
${\bf B}_{\mathrm{int}}$ is basically described by the ${\bf Q}$ = 0 propagation vector and that the in-plane components of 
${\bf B}_{\mathrm{int}}$ on the three sublattices are oriented at 120$^{\circ}$ to each other. 

Determination of the spin structure requires knowledge of the hyperfine 
coupling tensor ${\bf A}_{\mathrm{hf}}^{\mathrm{Cu}}$. As already mentioned, the unpaired $d$-electron 
should occupy the $d(3z^2 - r^2)$ orbital on the Cu sites with the $z$-direction making an angle of 26$^{\circ }$ 
with the kagome plane (Fig.~1). 
The hyperfine coupling for the $d(3z^2 - r^2)$ orbital has been well understood both theoretically and 
experimentally~\cite{Abragam,YoshidaCuCl}. It has an extremely large anisotropy with a
sign change due to an anisotropic dipolar field. The typical values are 
$A_{\mathrm{hf}}^{\mathrm{Cu}}(\|)$ = 10~T/$\mu _B$ for ${\bf M} \|  z$ and 
$A_{\mathrm{hf}}^{\mathrm{Cu}}(\bot)$ = $-$12~T/$\mu _B$ for ${\bf M} \bot z$. From the experimental value of
$B_{\mathrm{int}}$ = 7.7~T and the constraint $-$12 $\leq  A_{\mathrm{hf}} \leq $ 10~T/$\mu _B$, we conclude 
that the magnetic moment must be larger than 0.6~$\mu _B$ and its direction should be almost parallel or
perpendicular to the $z$-direction of the $d(3z^2 - r^2)$ orbital. Because of the
large anisotropy of ${\bf A}_{\mathrm{hf}}^{\mathrm{Cu}}$, the uniqueness of $B_{\mathrm{int}}$ requires 
the uniqueness of not only the magnitude of ${\bf M}$ but also the direction of ${\bf M}$ relative to ${\bf z}$. 
Since the  $d(3z^2 - r^2)$ orbitals and  ${\bf A}_{\mathrm{hf}}^{\mathrm{Cu}}$ approximately obey the 
symmetry of the isotropic kagome lattice, the spin structure should follow that of ${\bf B}_{\mathrm{int}}$.  
This leads to the ${\bf Q}$ = 0 and 120$^{\circ}$ spin structure shown in Fig.~1. 
Strictly speaking, however, the very small distortion in vesignieite should break the three fold symmetry, 
differentiating ${\bf B}_{\mathrm{int}}$ at the Cu1 and Cu2 sites. This may explain the fine structure 
and broadening of the Cu-NMR spectrum experimentally observed.  
 
Let us now examine if this spin structure is compatible with the $^{51}$V-NMR spectrum in a magnetic field. 
We again assume the symmetry of an isotropic kagome lattice. 
The hyperfine coupling between a V nucleus and the six nearest-neighbor Cu spins should be dominantly 
isotropic, but a small anisotropy (off-diagonal elements of the hyperfine coupling tensor) is expected. 
Since the sum of the in-plane components of the nearest-neighbor spins 
obeying the 120$^{\circ}$ structure cancels out, it makes no contribution to $B_{\mathrm{int}}$ at V nuclei 
if this anisotropy is neglected. 
However, when the small anisotropy is taken into account, the in-plane components can produce a finite $B_{\mathrm{int}}^{\mathrm{(1)}}$ 
perpendicular to the plane through the off-diagonal elements. 
We then expect a rectangular powder pattern with the width of 
$2B_{\mathrm{int}}^{\mathrm{(1)}}$. The observed $^{51}$V-NMR spectra at low temperatures (Fig.~3), 
however, show the superposition of two such powder patterns with different widths, 0.29 and 0.12~T. 
A simple possible method to account for the two values of $B_{\mathrm{int}}$ 
is the consideration of an out-of-plane spin component, which is uniform in the $ab$-plane but alternates along the $c^*$-direction. 
This component produces $B_{\mathrm{int}}^{\mathrm{(2)}}$ 
perpendicular to the plane with alternating signs through the main isotropic hyperfine coupling. 
The total internal field at the V sites
has two values, $|B_{\mathrm{int}}^{\mathrm{(1)}} + B_{\mathrm{int}}^{\mathrm{(2)}}|$ 
and $|B_{\mathrm{int}}^{\mathrm{(1)}} - B_{\mathrm{int}}^{\mathrm{(2)}}|$, leading to 
$B_{\mathrm{int}}^{\mathrm{(2)}}$ = 0.04 or 0.1 T. 
From the value $A_{\mathrm{hf}}^{\mathrm{V}}$ = 0.87 T/$\mu _B$, the out-of-plane component 
of ${\bf M}$ is estimated to be 0.05 or 0.12 $\mu _B$. Such a small modulation of the spin structure 
could also give an additional structure to the Cu-NMR spectrum observed in Fig. 5. 
It is noted that the antiferromagnetic out-of-plane moment 0.05 or 0.12 $\mu _B$ is much smaller than 
$M >$ 0.6 $\mu _B$, but is much larger than the ferromagnetic component $~0.01 \mu _B$ estimated from the magnetization.

The ${\bf Q}$ = 0 structure is favored in the kagome lattice with the DM interaction \cite{Cepas, Elhajal}. 
In fact, the large line width of the electron spin resonance in vesignieite~\cite{Zhang} indicates 
a large DM interaction~\cite{Zorko}. Therefore, vesignieite is likely to be an example of 
kagome antiferromagnets, in which a large DM interaction stabilizes the long-range magnetic order 
with ${\bf Q}$ = 0 and the 120$^{\circ}$ spin structure. 

Our conclusion is apparently inconsistent with the previous suggestion that 
small frozen moments coexist with dynamically fluctuating spins.~\cite{Colman,Quilliam} 
However, there are many similarities between the present and previous experimental results. 
For example, the transition temperature of 9 K is the same 
and the internal fields observed at the V sites in both experiments 
have similar magnitudes at the lowest temperature.~\cite{Quilliam} 
Therefore, both results seem to have the same physical origin. 
Although the reason for the large sample dependence is not well understood yet, 
we believe that our experiments on the high-quality sample capture the intrinsic properties of vesignieite. 

In summary, our measurements reveal a two-step 
magnetic transition in vesignieite: the appearance of spatially inhomogeneous static moments  
below 13 K and a long-range order below 9 K. The nearly unique internal field at Cu nuclei 
indicates that the spin structure is basically represented by a ${\bf Q}$ = 0 propagation vector 
and the 120$^{\circ}$ orientation of the sublattice magnetizations.  

\section*{Acknowledgment}
We thank H. Yoshida for valuable discussions. 
This work was supported by MEXT KAKENHI on Priority Areas 
``Novel State of Matter Induced by Frustration'' (No. 22014004), 
JSPS KAKENHI (B) (No. 21340093), and the MEXT-GCOE program.


\begin{thebibliography}{99} 
\bibitem{Waldtmann} C. Waldtmann, H.-U. Everts, B. Bernu, C. Lhuillier, 
P. Sindzingre, P. Lecheminant, and L. Pierre: Eur. Phys. J. B \textbf{2} (1998) 501. 
\bibitem{Hermele} M. Hermele, Y. Ran, P. A. Lee, and X.-G. Wen: Phys. Rev. B \textbf{77} (2008) 224413. 
\bibitem{Singh} R. R. P. Singh and D. A. Huse: Phys. Rev. B \textbf{76} (2007) 180407(R). 
\bibitem{Hiroi} Z. Hiroi, M. Hanawa, N. Kobayashi, M. Nohara, H. Takagi, 
Y. Kato, and M. Takigawa: J. Phys. Soc. Jpn. \textbf{70} (2001) 3377. 
\bibitem{HYoshida} H. Yoshida, Y. Okamoto, T. Tayama, T. Sakakibara, M. Tokunaga, A. Matsuo, Y. Narumi, 
K. Kindo, M. Yoshida, M. Takigawa, and Z. Hiroi: J. Phys. Soc. Jpn. \textbf{78} (2009) 043704. 
\bibitem{Yoshida1} M. Yoshida, M. Takigawa, H. Yoshida, Y. Okamoto, 
and Z. Hiroi: Phys. Rev. Lett. \textbf{103} (2009) 077207. 
\bibitem{Yoshida2} M. Yoshida, M. Takigawa, H. Yoshida, Y. Okamoto, 
and Z. Hiroi: Phys. Rev. B \textbf{84} (2011) 020410(R). 
\bibitem{YoshidaHF} M. Yoshida, M. Takigawa, S. Kr\"{a}mer, S. Mukhopadhyay, 
M. Horvati\'{c}, C. Berthire, H. Yoshida, Y. Okamoto, and Z. Hiroi: J. Phys. Soc. Jpn. \textbf{81} (2012) 024703. 
\bibitem{Okamoto} Y. Okamoto, H. Yoshida, and Z. Hiroi: J. Phys. Soc. Jpn. \textbf{78} (2009) 033701. 
\bibitem{Colman} R. H. Colman, F. Bert, D. Boldrin, A. D. Hillier, P. Manuel, 
P. Mendels, and A. S. Wills: Phys. Rev. B \textbf{83} (2011) 180416(R). 
\bibitem{Quilliam} J. A. Quilliam, F. Bert, R. H. Colman, D. Boldrin, 
A. S. Wills, and P. Mendels: Phys. Rev. B \textbf{84} (2011) 180401(R). 
\bibitem{HYoshida2} H. Yoshida, Y. Michiue, E. Takayama-Muromachi, and M. Isobe: J. Mater. Chem. \textbf{22} (2012) 18793. 
\bibitem{Shores} M. P. Shores, E. A. Nytko, B. M. Bartlett, and D. G. Nocera: J. Am. Chem. Soc. \textbf{127} (2005) 13462. 
\bibitem{Helton} J. S. Helton, K. Matan, M. P. Shores, E. A. Nytko, B. M. Bartlett, Y. Yoshida, 
Y. Takano, A. Suslov, Y. Qiu, J.-H. Chung, D. G. Nocera, and Y. S. Lee: Phys. Rev. Lett. \textbf{98} (2007) 107204. 
\bibitem{Cepas} O. C\'{e}pas, C. M. Fong, P. W. Leung, and C. Lhuillier: Phys. Rev. B \textbf{78} (2008) 140405(R). 
\bibitem{Schnyder} A. P. Schnyder, O. A. Starykh, and L. Balents: Phys. Rev. B \textbf{78} (2008) 174420. 
\bibitem{Wang} F. Wang, A. Vishwanath, and Y. B. Kim: Phys. Rev. B \textbf{76} (2007) 094421. 
\bibitem{Domenge} J.-C. Domenge, P. Sindzingre, C. Lhuillier, and L. Pierre: Phys. Rev. B \textbf{72} (2005) 024433. 
\bibitem{Reimers} J. N. Reimers and A. J. Berlinsky: Phys. Rev. B \textbf{48} (1993) 9539. 
\bibitem{Elhajal} M. Elhajal, B. Canals, and C. Lacroix: Phys. Rev. B \textbf{66} (2002) 014422. 
\bibitem{Zhesheng} M. Zhesheng, H. Ruilin, and Z. Xiaoling: Dizhi Xuebao \textbf{64} (1990) 302 [in Chinese]. 
\bibitem{Abragam} A. Abragam and B. Bleaney: {\it Electron Paramagnetic Resonance of Transition Ions} 
(Oxford Univ. Press, New York, 1980). 
\bibitem{YoshidaCuBr} M. Yoshida, N. Ogata, M. Takigawa, T. Kitano, H. Kageyama, 
Y. Ajiro, and K. Yoshimura: J. Phys. Soc. Jpn. \textbf{77} (2008) 104705. 
\bibitem{YoshidaCuCl} M. Yoshida, N. Ogata, M. Takigawa, J. Yamaura, M. Ichihara, T. Kitano, 
H. Kageyama, Y. Ajiro, and K. Yoshimura: J. Phys. Soc. Jpn. \textbf{76} (2007) 104703. 
\bibitem{Zhang} W.-M. Zhang, H. Ohta, S. Okubo, M. Fujisawa, T. Sakurai, Y. Okamoto, 
H. Yoshida, and Z. Hiroi: J. Phys. Soc. Jpn. \textbf{79} (2010) 023708. 
\bibitem{Zorko} A. Zorko, S. Nellutla, J. van Tol, L. C. Brunel, F. Bert, F. Duc, J.-C. Trombe, 
M. A. de Vries, A. Harrison, and P. Mendels: Phys. Rev. Lett. \textbf{101} (2008) 026405. 

\end{thebibliography}
\end{document}